\documentclass[10pt, conference]{IEEEtran}
\IEEEoverridecommandlockouts
\usepackage{amsmath,amsfonts}
\usepackage{algorithm}
\usepackage{algpseudocode}

\usepackage{array}
\usepackage{textcomp}
\usepackage{stfloats}
\usepackage{verbatim}
\usepackage{graphicx}
\usepackage{cite}
\usepackage{multirow}
\usepackage{subcaption} 
\usepackage{caption}    
\usepackage{float}
\usepackage{placeins}
\usepackage{wrapfig}
\usepackage[hidelinks,colorlinks=true,linkcolor=black,citecolor=black,urlcolor=black]{hyperref}
\usepackage{makecell}
\usepackage{tikz}

\def\BibTeX{{\rm B\kern-.05em{\sc i\kern-.025em b}\kern-.08em
    T\kern-.1667em\lower.7ex\hbox{E}\kern-.125emX}}

\usepackage{titlesec}
\titlespacing*{\section}{0pt}{*0.8}{*0.8}
\titlespacing*{\subsection}{0pt}{*0.6}{*0.6}
\titlespacing*{\subsubsection}{0pt}{*0.5}{*0.5}

\begin{document}

\title{Network Anomaly Detection in Distributed Edge Computing Infrastructure}
\author{\IEEEauthorblockN{William Marfo, Enrique A. Rico, Deepak K. Tosh, Shirley V. Moore} \\
\IEEEauthorblockA{\textit{Department of Computer Science}, \textit{University of Texas at El Paso}, El Paso, USA \\
\{wmarfo, earico\}@miners.utep.edu, dktosh@utep.edu, svmoore@utep.edu
}}

\maketitle

\begin{abstract}
As networks continue to grow in complexity and scale, detecting anomalies has become increasingly challenging, particularly in diverse and geographically dispersed environments. Traditional approaches often struggle with managing the computational burden associated with analyzing large-scale network traffic to identify anomalies. This paper introduces a distributed edge computing framework that integrates federated learning with \textit{Apache Spark} and Kubernetes to address these challenges. We hypothesize that our approach, which enables collaborative model training across distributed nodes, significantly enhances the detection accuracy of network anomalies across different network types. We show that by leveraging distributed computing and containerization technologies, our framework not only improves scalability and fault tolerance but also achieves superior detection performance compared to state-of-the-art methods. Extensive experiments on the UNSW-NB15 and ROAD datasets validate the effectiveness of our approach, demonstrating statistically significant improvements in detection accuracy and training efficiency over baseline models, as confirmed by Mann-Whitney \(\mathrm{U}\) and Kolmogorov-Smirnov tests $(p < 0.05)$.
\end{abstract}

\footnote{This material is based upon work supported by the United States Department of Energy’s (DOE) Office of Fossil Energy (FE) Award DE-FE0031744.}


\section{Introduction} 
The rapid growth of digital connectivity and internet adoption have revolutionized communication and interaction worldwide. Networks are at the core of this digital ecosystem, facilitating seamless data transmission across vast distances. However, as networks become increasingly complex and interconnected, detecting network anomalies has become a critical challenge \cite{fed,marfo2024detectingmasqueradeattackscontroller}. Traditional methods such as rule-based or signature-based techniques often fall short, particularly when faced with emerging or previously unseen threats \cite{18,b1}. Furthermore, the substantial computational workload required to process and analyze large-scale network traffic data compounds the complexity of the problem \cite{18}. Centralized machine learning (ML) approaches, which gather sensitive network data on central servers, exacerbate concerns about data breaches, privacy violations, and cross-border data security \cite{comparison1, client}.

The challenge of detecting network anomalies extends beyond traditional IT networks to diverse domains, including the Internet of Things (IoT) and automotive systems. These networks, such as those in modern vehicles, are particularly susceptible to sophisticated attacks like masquerade attacks, which mimic legitimate communication to manipulate system behavior without immediate detection \cite{marfo2024detectingmasqueradeattackscontroller}. Traditional intrusion detection systems (IDS) often struggle with such stealthy attacks, especially in handling the high-dimensional data typical of diverse network environments. Our work addresses this gap by applying our federated learning (FL) framework to both general network traffic (UNSW-NB15 dataset) \cite{b1} and controller area network traffic (ROAD dataset) \cite{verma2022addressing}, demonstrating its versatility and effectiveness across various network types. We chose the UNSW-NB15 dataset because it comprehensively represents modern network traffic patterns and includes diverse attack types. To complement this, we selected the ROAD dataset for its unique focus on automotive network traffic, especially its realistic masquerade attacks. By using these datasets, we validate our approach across both general and specialized network environments, enhancing the detection of general network anomalies while also demonstrating promise in identifying subtle, sophisticated attacks in specialized network protocols. This comprehensive validation contributes to improved security across a wide range of connected systems. 

Motivated by these challenges and the limitations of existing approaches, we propose a framework that integrates FL with \textit{Apache Spark} \cite{19} and Kubernetes \cite{kub}, enabling scalable, efficient, and privacy-preserving network anomaly detection across diverse and geographically dispersed environments. This approach facilitates collaborative model training without centralizing sensitive data, thus enhancing privacy and security. Our framework effectively addresses the challenges of maintaining model accuracy across diverse network environments and the computational demands of processing large-scale network traffic data. By leveraging distributed computing and containerization technologies, we achieve improved scalability and fault tolerance, crucial for the real-world deployment of network anomaly detection systems. The primary contributions of this paper are:
\begin{itemize}
\item We introduce a distributed edge computing architecture that integrates FL with \textit{Apache Spark} and Kubernetes for efficient and scalable network anomaly detection, demonstrating improved detection accuracy and training efficiency compared to baseline methods.

\item We develop an adaptive checkpointing mechanism using Weibull distribution modeling that enhances fault tolerance, enabling robust performance as the number of clients increases and under various dropout scenarios.

\item We validate our framework's effectiveness across diverse network environments by applying it to both the UNSW-NB15 and ROAD datasets, showcasing its capability in detecting general network anomalies and specialized automotive cybersecurity threats.
\end{itemize}

Our experiments demonstrate improved detection accuracy (97.5\% on UNSW-NB15, 91.4\% on ROAD) and training efficiency over current methods. The paper structure includes background (\S II), related work (\S III), FL framework (\S IV), evaluation (\S V), and conclusion (\S VI).

\section{Background}

This section outlines the key concepts of FL and its role in enhancing network anomaly detection, providing context for the proposed framework.

\subsection{Federated Learning}
FL enables collaborative model training across multiple edge devices without centralizing data, thereby preserving privacy and reducing communication costs \cite{fed, comparison1}. A prominent FL framework is federated averaging (FedAvg) \cite{new33}, where updates from selected clients are averaged to update a global model, achieving reliable convergence. Managing distributed learning in large-scale environments requires addressing challenges in scalability and fault tolerance. \textit{Apache Spark} \cite{19} plays a crucial role in handling large datasets with its in-memory processing capabilities, while Kubernetes \cite{kub} together with Docker \cite{15} provide infrastructure for scaling and managing distributed applications across clusters. The integration of \textit{Spark} with Kubernetes optimizes submission processes and reduces time for iterative algorithms used in distributed ML.

\subsection{Application in Network Anomaly Detection}
Network anomaly detection involves identifying unusual patterns in network traffic that may indicate security threats. FL is well-suited for this domain, enabling models to be trained on distributed data while preserving privacy and supporting real-time detection. Prior studies have demonstrated FL’s potential in improving network security, particularly in intrusion detection on edge nodes and cloud servers \cite{15, fed, 18}. However, scaling FL for network anomaly detection still presents challenges, especially in managing computational demands and ensuring fault tolerance. Our framework addresses these challenges by integrating FL with \textit{Apache Spark} and Kubernetes to create a scalable, efficient distributed edge computing environment capable of handling large-scale network traffic data while maintaining privacy.

\section{Related Work}

Diro et al. \cite{18} proposed a distributed deep learning-based IoT/Fog network attack detection system, demonstrating superior performance over centralized systems, particularly in detecting small mutations due to deep models' high-level feature extraction capabilities. Lui et al. \cite{client} introduced a client-edge-cloud hierarchical FL architecture with the HierFAVG algorithm, reducing model training time and energy consumption compared to traditional cloud-based FL. Kim et al. \cite{comparison1} proposed an FL-based collaborative anomaly detection system with multiple edge nodes and a server, preserving user privacy. Sáez-de-Cámara et al. \cite{saez2023clustered} proposed an FL-based architecture with an unsupervised clustering algorithm for network intrusion detection in large IoT deployments, achieving faster convergence and improved attack detection. Julian et al. Jullian et al. \cite{jullian2023deep} implemented a distributed deep learning framework with an LSTM model to enhance detection accuracy of malicious traffic in IoT networks.

Compared to previous studies, this work uniquely integrates FL with \textit{Apache Spark} and Kubernetes for network anomaly detection in distributed edge computing. We introduce an adaptive checkpointing mechanism using Weibull distribution modeling \cite{vardhan2024modern}, enhancing fault tolerance. Our approach demonstrates improved accuracy and efficiency on both general (UNSW-NB15) and automotive (ROAD) network datasets. The framework maintains high performance as client numbers increase and shows resilience against dropouts, addressing scalability challenges not fully explored in existing literature. This comprehensive solution offers robust anomaly detection in complex, distributed environments.

\section{Proposed Federated Learning Framework for Network Anomaly Detection}
We present a FL framework for network anomaly detection that leverages distributed edge computing to enhance detection accuracy and efficiency. Our approach integrates FL with \textit{Apache Spark} \cite{19} and Kubernetes \cite{kub} to create a scalable, fault-tolerant system capable of efficiently processing large-scale network traffic data.

\subsection{System Architecture}
Our architecture integrates three key components: a distributed learning framework, a distributed data processing engine, and a container orchestration platform. This combination provides a scalable, fault-tolerant solution tailored for network anomaly detection, ensuring efficient management of computational workloads across distributed nodes.
The distributed learning framework coordinates model training across multiple clients, enabling collaborative learning without centralizing raw data. The distributed data processing engine facilitates the efficient handling of large-scale network traffic data, significantly reducing the time required for iterative algorithms common in ML tasks.
Our container orchestration platform manages deployment, scaling, and resource allocation across a cluster of nodes, enhancing the system's adaptability to varying network sizes and data volumes. The cluster consists of a master node and multiple executor nodes, which facilitate the parallel processing of tasks.

\subsection{Federated Learning Architecture for Network Anomaly Detection}

\begin{figure}[h]
\centerline{\includegraphics[width=0.42\textwidth]{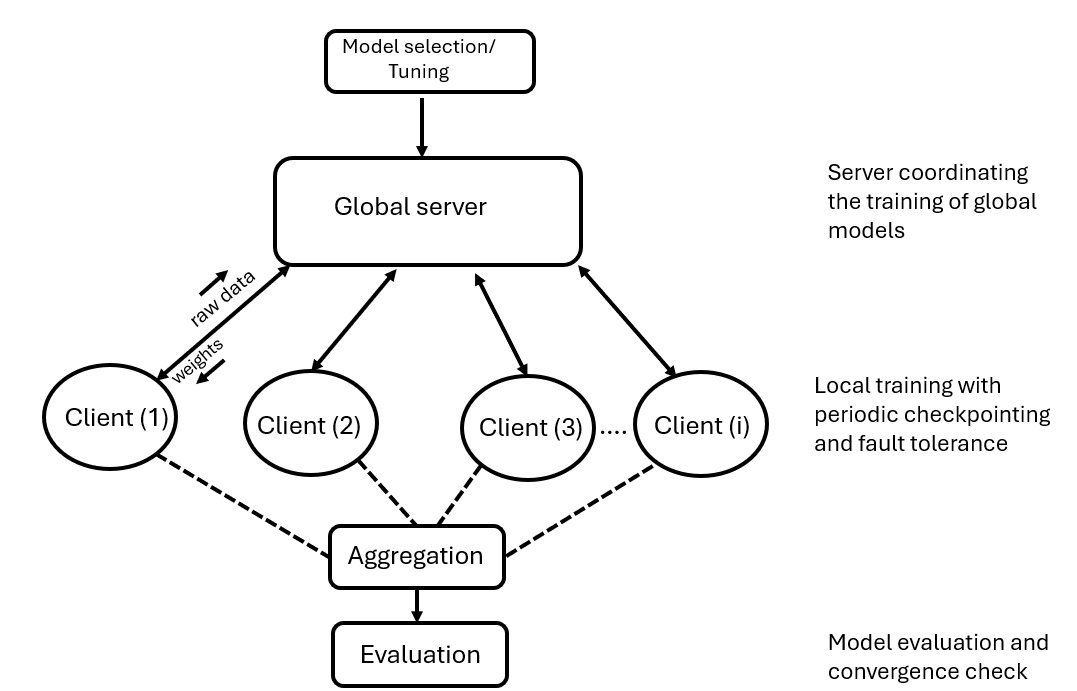}}
\caption{FL architecture for network anomaly detection, illustrating model selection, training, checkpointing, aggregation, and evaluation.}
\label{fig1}
\end{figure}

Our FL environment for network anomaly detection consists of two main components: \textit{clients} and a \textit{global server}. Fig.~\ref{fig1} illustrates the overall architecture of our FL system.

\textbf{1. Client:}
A client is a device or machine that owns the network traffic data. To preserve privacy, each client’s data remains local and is not shared directly with other clients or the global server. Each client maintains a local model, which is an independent copy of the global deep learning model for anomaly detection.
The local model on each client is trained for a few epochs on the client's local data. Let us assume we have $n$ number of clients symbolized as $c_i$, where $i \in {1, \ldots, n}$. Each client has its data $X_i$ and a local model $f_i$, where $X_i \in \mathbf{R}^{m_i\times d}$, $m_i$ is the number of samples for client $i$, and $d$ is the number of features in each sample. After training $f_i$ on $e$ epochs on $X_i$ data, we pass updated parameters $w_{f_i}$ of local $f_i$ model to the global server. This can be represented as $w_{f_i} = f_i(X_i, e)$.

\smallskip
\textbf{2. Global server:}
The global server hosts the global model, and the parameters of this model are relayed to all clients after performance evaluation. The global server aggregates parameters received from all clients based on an aggregation function. Assuming the global server \(g\) is connected to \(n\) clients, it aggregates the parameters received from these clients as \(w_g = \frac{1}{n} \sum_{i=1}^n w_{f_i}\) and updates its global model \(h\) accordingly.

\vspace{-1em}\vspace{-1em}

This iterative process allows the model to learn from diverse network environments without centralizing sensitive data. The integration of \textit{Apache Spark} and Kubernetes enhances the scalability and efficiency of this process, enabling seamless management of resources and distributed computation between the \textit{global server} and \textit{clients}. The FL process allows models to be trained on diverse and distributed datasets that reflect various network conditions and behaviors. By aggregating knowledge from multiple geographically dispersed clients, the global model gains a more comprehensive understanding of network traffic patterns, enhancing its ability to detect anomalies that may be specific to certain environments or conditions. This distributed approach also ensures that models are continuously updated and improved, adapting to new threats emerging across different networks. The FL process for network anomaly detection is outlined in Algorithm~\ref{alg:fl_anomaly}. Checkpointing details and related notations are covered in §IV-B.

\begin{algorithm}
\caption{FL training with fault tolerance and checkpointing}
\label{alg:fl_anomaly}
\begin{algorithmic}[1]
\Require Training data \(X_i\) for each client \(i\), early stopping patience \(p\), optimal checkpointing interval \(t_c^*\)
\Ensure Global model \(w_g\)
\State Initialize \(w_g\), \(f_i\) for each client, best\_performance \(\leftarrow -\infty\), patience\_counter \(\leftarrow 0\)
\For{round \(r = 0\) to \(\text{max\_rounds}\)}
    \For{each client \(i\) in parallel}
        \State \(f_i \leftarrow w_g\)
        \State last\_checkpoint \(\leftarrow\) current\_time()
        \For{epoch = 1 to \(\text{num\_epochs}\)}
            \State Train \(f_i\) on \(X_i\)
            \If{current\_time() - last\_checkpoint \(\geq t_c^*\)}
                \State Checkpoint \(f_i\), optimizer state
                \State last\_checkpoint \(\leftarrow\) current\_time()
            \EndIf
            \If{client\_failure\_detected()}
                \State Recover from checkpoint or reinitialize with \(w_g\)
            \EndIf
        \EndFor
        \State Send \(w_{f_i}\) to global server
    \EndFor
    \State \(w_g \leftarrow \frac{1}{n} \sum_{i=1}^n w_{f_i}\)
    \State performance \(\leftarrow\) Evaluate(\(w_g\))
    \If{performance improves}
        \State best\_performance \(\leftarrow\) performance, patience\_counter \(\leftarrow 0\)
    \Else
        \State patience\_counter \(\leftarrow\) patience\_counter \(+ 1\)
    \EndIf
    \If{patience\_counter \(\geq p\)}
        \State \textbf{break}
    \EndIf
\EndFor
\State \Return \(w_g\)
\end{algorithmic}
\end{algorithm}

\vspace{-1em}\textbf{Termination condition/epochs in Algorithm~\ref{alg:fl_anomaly}:} The training process in this algorithm uses a combination of iterative rounds (denoted by \texttt{max\_rounds}) and an early stopping mechanism. Training continues as long as performance improves. If no improvement is observed for \(p\) consecutive rounds, as indicated by the \texttt{patience\_counter}, training halts. Each client trains for a fixed number of local epochs (\texttt{num\_epochs}) per round. The checkpointing mechanism, triggered at intervals of \(t_c^*\), ensures quick recovery from interruptions and maintains continuity even with client dropouts or failures. The algorithm also includes a client failure detection and recovery mechanism, which, combined with checkpointing, sustains training continuity even under adverse conditions.

\textbf{Handling training failures in decentralized ML:}

In FL for network anomaly detection within distributed edge environments, fault tolerance is crucial due to potential client dropouts or disconnections \cite{benoit2024checkpointing}. Our framework incorporates a checkpointing mechanism to ensure smooth recovery in case of client failures.

\paragraph{Recovery protocol without checkpointing}
Without checkpointing, recovery can proceed by either restarting the entire training process or re-initializing the failed client’s model with the most recent global weights \cite{benoit2024checkpointing}. We favor the latter as it minimizes disruption and maintains overall training progress, though with a slight risk of temporary inconsistencies.

\paragraph{Recovery protocol with checkpointing}
With checkpointing, each client regularly saves its model state as binary files \cite{benoit2024checkpointing}. If a failure occurs, the system restores the client's state from the last checkpoint, allowing training to resume without starting over. If a failure occurs during aggregation, the global server either waits for recovery or redistributes the client’s data to other active clients, ensuring training continuity.

\paragraph{Optimal checkpointing interval}
We model the likelihood of client failure using a Weibull distribution \cite{vardhan2024modern}, which is effective for distributed systems. The probability of failure within a checkpointing interval \( t_c \) is \( p_f(t_c) = 1 - \exp\left(-\left(\frac{t_c}{\lambda}\right)^k\right) \), where \( \lambda \) and \( k \) are scale and shape parameters. The cost function balancing checkpointing overhead with recovery costs is \( C(t_c) = \frac{t_c}{T} + p_f(t_c) \cdot \frac{t_r}{T} \), where \( T \) is total computation time and \( t_r \) is recovery time. The optimal checkpointing interval \( t_c^* \) is determined by solving \( \frac{dC}{dt_c} = 0 \) numerically, based on estimated \( \lambda \) and \( k \) from historical failure data.

\section{Evaluation Results}

\subsection{Experimental Setup}
Experiments were conducted on a system with an Intel\textsuperscript{\textregistered} Core\textsuperscript{\texttrademark} i9-12900HK CPU, NVIDIA GeForce RTX 3080 Ti GPU, and 32GB RAM. The implementation used Python 3.8.18, with \texttt{TensorFlow 2.6.0} for model training, \texttt{PyTorch 0.5.0} for the FL framework, \texttt{Apache Spark 3.1.2} for distributed processing, \texttt{Kubernetes Python Client 28.1.0} for cluster management, and \texttt{scikit-learn 0.24.2} for evaluation metrics. The FL environment was deployed on a Kubernetes cluster with one master node and five worker nodes, each running Spark executors within Kubernetes pods.

\subsubsection{Datasets}
Our study utilizes two datasets to evaluate the performance of our FL framework for network anomaly detection:

\paragraph{UNSW-NB15 dataset} 
The UNSW-NB15 dataset \cite{b1} is a comprehensive network intrusion dataset developed at the UNSW Cybersecurity Lab in Canberra, Australia. It captures the complexities of modern network traffic scenarios, including a wide range of low-footprint intrusions. The dataset was generated using the IXIA PerfectStorm tool, resulting in a balanced mix of genuine contemporary standard activities and recent synthetic attack behaviors.
The dataset comprises 2,540,043 samples, each with 49 features that capture various aspects of network packets. These features were extracted using Argus, Bro-IDS tools, and twelve distinct algorithms. Each sample is labeled binary, where `1' indicates an attack or anomaly, and `0' represents normal traffic. Table \ref{table:classdist} details the class distribution of the UNSW-NB15 dataset.

\begin{table}[h]
\centering
\caption{Class distribution of the UNSW-NB15 dataset}
\label{table:classdist}
\begin{tabular}{l|r|r}
\hline
\textbf{Category} & \textbf{Training Set} & \textbf{Testing Set} \\ \hline
Normal            & 56,000                & 37,000               \\
Generic           & 40,000                & 18,871               \\
Exploits          & 33,393                & 11,132               \\
Fuzzers           & 18,184                & 6,062                \\
DoS               & 12,264                & 4,089                \\
Reconnaissance    & 10,491                & 3,496                \\
Analysis          & 2,000                 & 677                  \\
Backdoor          & 1,746                 & 583                  \\
Shellcode         & 1,133                 & 378                  \\
Worms             & 130                   & 44                   \\
\hline
\textbf{Total}    & 175,341               & 82,332               \\
\hline
\end{tabular}
\end{table}

\vspace{-1em}\paragraph{ROAD dataset}
We also evaluate our framework on the Real ORNL Automotive Dynamometer (ROAD) dataset \cite{verma2022addressing}, which contains controller area network (CAN) data collected from a real vehicle at Oak Ridge National Laboratory. This dataset is particularly valuable for its inclusion of physically verified fabrication and simulated masquerade attacks, providing a realistic environment for testing CAN security methods.
The ROAD dataset comprises 3.5 hours of recorded data, with 3 hours used for training and 30 minutes for testing. While the dataset includes various types of masquerade attacks, our study focuses specifically on the correlated signal masquerade attack, which injects varying values for wheel speeds, resulting in the vehicle coming to a halt.

\subsubsection{Data preprocessing}
For the UNSW-NB15 dataset, we performed several preprocessing steps to prepare the data for our FL model. Initially comprising 49 features, we removed irrelevant columns and addressed mixed data types. Categorical features such as protocol type and connection status were encoded using one-hot encoding. Numerical features were normalized to zero mean and unit variance to ensure equal feature importance. IP addresses were mapped to unique identifiers to facilitate efficient processing in our distributed environment.
For the ROAD dataset, focusing on the correlated signal masquerade attack, we followed the preprocessing steps outlined in \cite{marfo2024detectingmasqueradeattackscontroller}.

\subsubsection{Model Architecture} 
Our deep neural network model is designed for binary network anomaly detection. It takes 43 relevant network traffic features as input and outputs a probability value via a sigmoid function. The model is trained using binary cross-entropy loss and the Adam optimizer with an adaptive learning rate starting at 0.001. The architecture comprises dense layers with 1024, 768, 512, 256, 128, 64, and 32 neurons, each with ReLU activations. Batch normalization layers are interspersed for training stability, and dropout layers are paired with the last three dense layers to prevent overfitting. The network culminates in an output layer with a single neuron and sigmoid activation.

\subsubsection{Performance Metrics}
\label{sec:PerformanceMetrics}

To evaluate the effectiveness of our framework, we use key performance metrics that provide comprehensive insights for detecting network anomaly anomalies.

\textit{(i) Accuracy} measures the proportion of correct predictions among all instances, offering a broad view of model performance. However, it may not fully capture effectiveness in imbalanced datasets where anomalies are much less frequent than normal instances.

\textit{(ii) AUC-ROC} (Area under the receiver operating characteristic curve) evaluates the model's ability to discriminate between classes across different thresholds. It is defined as the integral of the true positive rate (TPR) against the false positive rate (FPR), mathematically expressed as $\text{AUC-ROC} = \int_{0}^{1} \text{TPR}(\text{FPR}^{-1}(x)) \, dx$. AUC-ROC values range from 0 to 1, with values closer to 1 indicating better performance, meaning the model is more effective at distinguishing between positive and negative classes. A value of 0.5 suggests no discriminative power, while values below 0.5 indicate performance worse than random guessing.

\subsubsection{Baselines}
To demonstrate the effectiveness of the proposed algorithm, we compare it against the following baselines in literature using 6 clients and a global server: (1) \textbf{FedAvg}\cite{new33}, where the global model is updated only after receiving updates from all clients; and (2) \textbf{FedL2P}\cite{NEURIPS2023_2fb57276}, which employs a meta-learning approach to optimize hyperparameters for personalized fine-tuning under data heterogeneity by learning a meta-network that outputs near-optimal hyperparameters based on client data profiles.

\subsection{Results and Analysis}

\subsubsection{Detection performance evaluation}
Table \ref{tab:perf_comparison} compares the performance of FedAvg, FedL2P, and our proposed method on the UNSW-NB15 and ROAD datasets over 300 epochs. Our method consistently outperforms the baselines in accuracy, AUC-ROC scores, and training time due to its efficient integration of FL with \textit{Apache Spark}  and Kubernetes and its adaptive checkpointing mechanism. On UNSW-NB15, we achieved 97.5\% accuracy with a 300-second training time, while on ROAD, we reached 91.4\% accuracy and 0.89 AUC-ROC in 430 seconds.

\begin{figure}[h]
\centering
\includegraphics[width=0.50\textwidth]{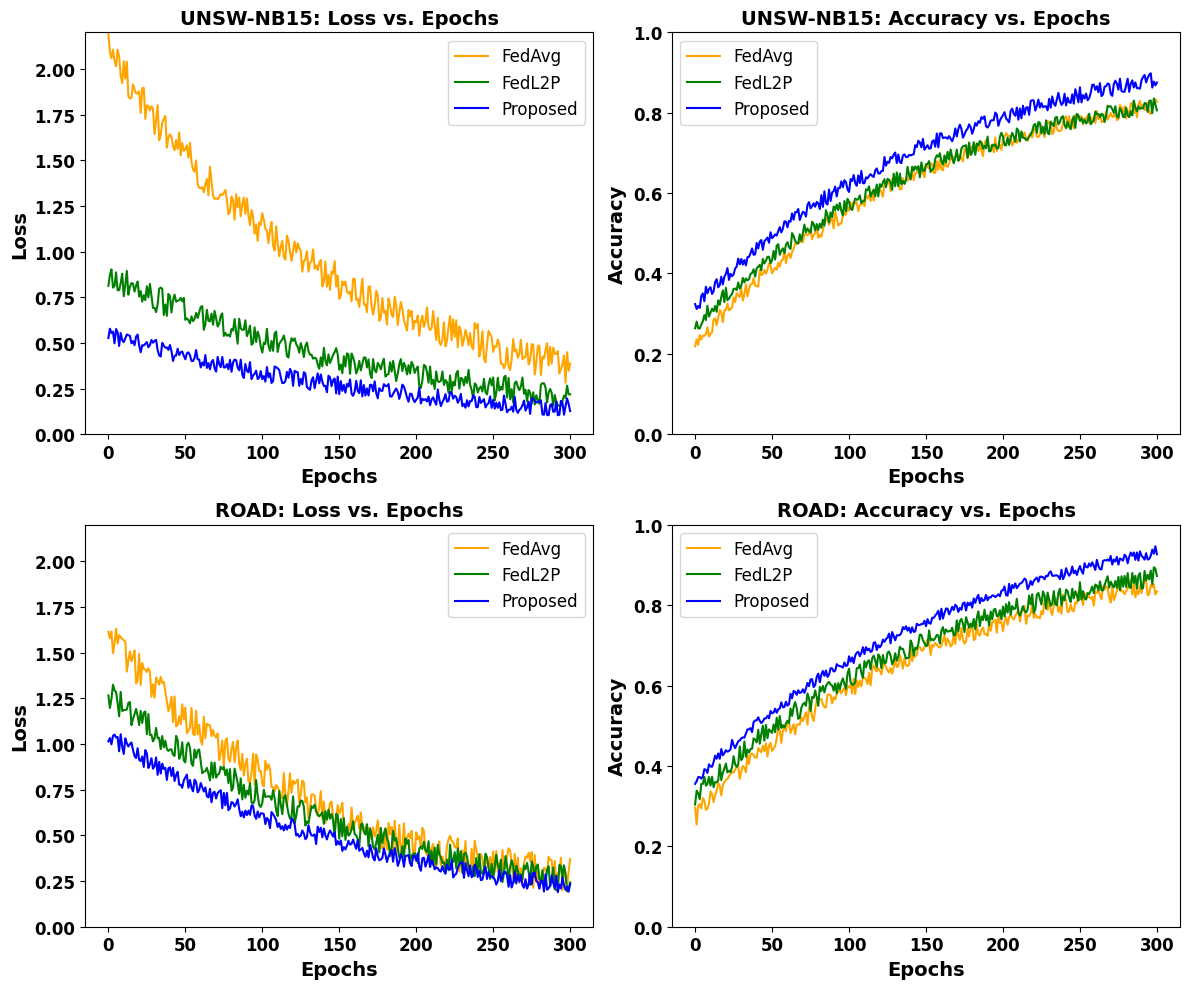}
\caption{Training performance of models in terms of loss and accuracy over 300 epochs on the UNSW-NB15 and ROAD datasets.}
\label{fig}
\end{figure}

Fig.~\ref{fig} illustrates the training performance in terms of loss and accuracy. Our method's faster convergence and higher stability, evident in both datasets, stem from improved model aggregation and efficient handling of client heterogeneity. On UNSW-NB15, our method stabilizes at \(\approx 0.97\) accuracy, compared to \(\approx 0.90\) for FedL2P and \(\approx 0.89\) for FedAvg. Similarly, on ROAD, we achieve \(\approx 0.91\) peak accuracy, while FedL2P and FedAvg reach \(\approx 0.89\) and \(\approx 0.85\), respectively.

\begin{table}[h]
\centering
\caption{Performance comparison of FedAvg, FedL2P, and Proposed method.}
\label{tab:perf_comparison}
\begin{tabular}{l|c|c|c}
\hline
\textbf{Method} & \textbf{Accuracy (\%)} & \textbf{AUC-ROC} & \textbf{Time (s)} \\ \hline
\multicolumn{4}{c}{\textbf{UNSW-NB15}} \\ \hline
FedAvg          & 0.89                   & 0.88             & 600              \\
FedL2P          & 92.1                   & 0.91             & 550              \\
Proposed        & \textbf{97.5}          & \textbf{0.96}    & \textbf{300}     \\ \hline
\multicolumn{4}{c}{\textbf{ROAD}} \\ \hline
FedAvg          & 85.3                   & 0.82             & 720              \\
FedL2P          & 88.7                   & 0.86             & 670              \\
Proposed        & \textbf{91.4}          & \textbf{0.89}    & \textbf{430}     \\
\hline
\end{tabular}\vspace{-1em}
\end{table}

\subsubsection{Scalability and fault tolerance analysis}

\begin{figure}[h]
\centering
\includegraphics[width=0.50\textwidth]{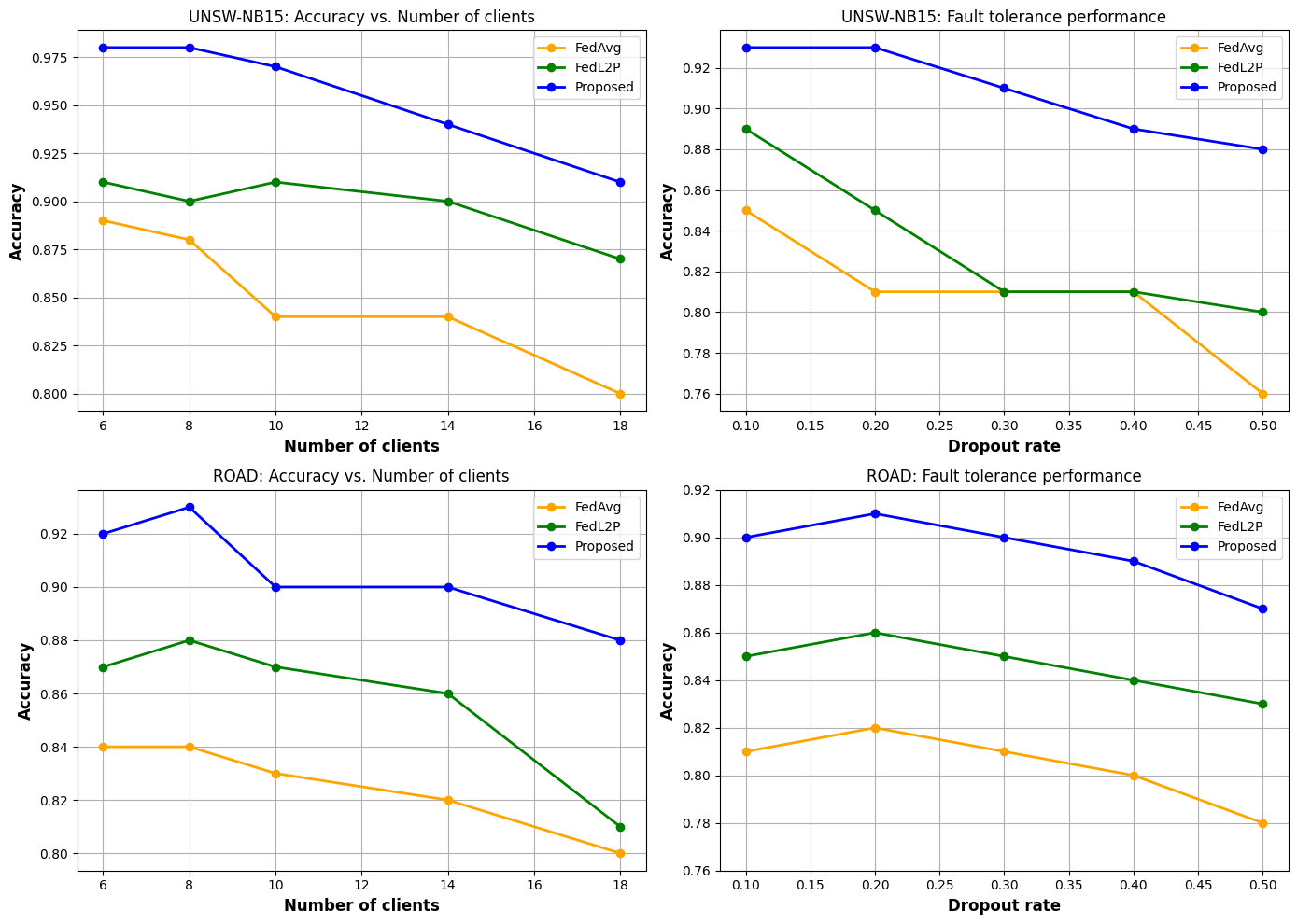}
\caption{Performance comparison on the UNSW-NB15 dataset (top row) and the ROAD dataset (bottom row). The left column shows accuracy as a function of the number of clients, and the right column shows accuracy under different dropout rates.}
\label{fig:scalability_fault}\vspace{-1em}
\end{figure}

We evaluate the scalability of our proposed method by observing accuracy trends as the number of clients increases. Fig.~\ref{fig:scalability_fault} shows that our proposed method consistently outperforms FedAvg and FedL2P across both UNSW-NB15 and ROAD datasets, maintaining high accuracy even as client numbers grow. Although the accuracy plateaus and slightly decreases beyond a certain number of clients, this can be attributed to the increased communication overhead and model aggregation complexity. Despite this, our method’s superior aggregation strategy still ensures better overall performance compared to the baselines.

For fault tolerance, we simulated client dropouts at varying rates. Fig.~\ref{fig:scalability_fault} shows that our approach exhibits a more gradual decline in accuracy compared to the baselines, demonstrating enhanced robustness against client failures. This resilience is largely due to our robust checkpointing mechanism, which periodically saves the model’s state during training. When a client drops out, tasks are quickly reassigned, and training resumes from the last checkpoint, minimizing the impact on performance. 

It is worth noting that in FL literature, trade-offs have been reported between communication cost and training time depending on the communication frequency with servers \cite{fed}. While our results demonstrate improved efficiency, a comprehensive investigation of bandwidth usage versus accuracy/efficiency trade-offs in our framework is delegated to future work.

\subsubsection{Statistical significance testing}

To further validate the differences in detection effectiveness among the methods (FedAvg, FedL2P, and our proposed framework), we employed the Mann-Whitney \(\mathrm{U}\) \cite{mann1947test} and the Kolmogorov-Smirnov \((\mathrm{KS}\)) \cite{berger2014kolmogorov} tests. The Mann-Whitney \(\mathrm{U}\) test \cite{mann1947test} is a non-parametric statistical test that evaluates whether there is a significant difference between two independent samples. The \(\mathrm{KS}\) test \cite{berger2014kolmogorov} is another non-parametric test that assesses whether two samples come from the same distribution.

In our analysis, we compared the AUC-ROC value distributions of the three methods across the two datasets (UNSW-NB15 and ROAD). The null hypothesis is that the AUC-ROC values for the proposed method are less than or equal to those for FedAvg and FedL2P, while the alternative hypothesis is that the AUC-ROC values for the proposed method are greater than those for the baselines. A low \( p \)-value indicates a significant difference between these methods. We focused on a significance level of \(\alpha = 0.05\). Table \ref{tab:test_results} presents the results of the Mann-Whitney \(\mathrm{U}\) and \(\mathrm{KS}\) tests for both datasets. In all cases, the tests yielded low \( p \)-values, indicating significant deviations from the expected distributions and supporting our hypothesis that the proposed method outperforms the baselines. Consequently, we reject the null hypothesis, confirming that the proposed method performs significantly better regarding AUC-ROC.

\begin{table}[h]
\centering
\caption{Mann-Whitney \(\mathrm{U}\) and \(\mathrm{KS}\) test results for AUC-ROC comparisons across methods.}
\begin{tabular}{|p{1.7cm}|c|c|c|c|}
\hline
\textbf{Dataset} & \multicolumn{2}{c|}{\textbf{Mann-Whitney \(\mathrm{U}\)}} & \multicolumn{2}{c|}{\textbf{Kolmogorov-Smirnov \(\mathrm{KS}\)}} \\
\cline{2-5}
 & \textbf{U Statistic} & \textbf{P-value} & \textbf{Statistic} & \textbf{P-value} \\
\hline
UNSW-NB15 & 10234.0 & 3.45e-15 & 0.471 & 2.93e-11 \\
ROAD      & 9785.0  & 1.02e-08 & 0.359 & 1.25e-07 \\
\hline
\end{tabular}
\label{tab:test_results}
\end{table}

\section{Conclusion}
This paper presents a distributed edge computing FL framework for network anomaly detection that outperforms FedAvg and FedL2P on the UNSW-NB15 and ROAD datasets. Our method achieved higher accuracy (97.5\% on UNSW-NB15, 91.4\% on ROAD) and AUC-ROC scores with reduced training time, confirmed by statistical tests. The framework demonstrated enhanced scalability and fault tolerance under increasing client numbers and dropouts. However, challenges may arise in extremely heterogeneous or imbalanced datasets. Future work will focus on integrating more complex anomaly detection algorithms and exploring applications in emerging technologies like 6G and cyberphysical systems.

\end{document}